\documentclass[nofootinbib,twocolumn,aps,pre,superscriptaddress,citeautoscript,floatfix]{revtex4-1}
\usepackage{graphics}
\usepackage{graphicx}
\usepackage{mathrsfs}

\begin{document}

\title{Defect-free Perpendicular Diblock Copolymer Films:\\
The Synergistic Effect of Surface  Topography and Chemistry}
\author{Xingkun Man$^{*}$}
\affiliation{Center of Soft Matter Physics and its Applications, Beihang University, Beijing 100191, China}
\affiliation{School of Physics and Nuclear Energy Engineering, Beihang University, Beijing 100191, China}
\author{Pan Zhou}
\affiliation{Department of Physics, Beijing Normal University, Beijing 100875, China}
\author{Jiuzhou Tang}
\affiliation{Beijing National Laboratory Molecular Sciences, Joint Laboratory of Polymer Science and Materials,
Institute of Chemistry, Chinese Academy of Sciences, Beijing 100190, China}
\author{Dadong Yan}
\affiliation{Department of Physics, Beijing Normal University, Beijing 100875, China}
\author{David Andelman$^{*,}$}
\affiliation{Raymond and Beverly Sackler School of Physics and Astronomy, Tel Aviv University, Ramat Aviv 69978, Tel Aviv, Israel}
\affiliation{CAS Key Laboratory of Theoretical Physics, Institute of Theoretical Physics, Chinese
Academy of Sciences,
Beijing 100190, China}
\date{Aug 19, 2016}

\begin{abstract}

We propose a direct self-assembly mechanism towards obtaining defect-free perpendicular lamellar phases of diblock copolymer (BCP) thin films.
In our numerical study, a thin BCP film having a flat top surface is casted on a uni-directional corrugated solid substrate. The substrate is treated chemically and has a weak preference toward one of the two BCP components. Employing self-consistent field theory (SCFT), we find that there is an enhanced synergy between two substrate characteristics: its topography (geometrical roughness) combined with a weak surface preference. This synergy produces the desired perpendicular lamellar phase with perfect inplane ordering. Defect-free BCP lamellar phases are reproducible for several random initial states, and  are obtained for a range of substrate roughness and chemical characteristics, even for a uni-directional multi-mode substrate roughness. Our theoretical study suggests possible experiments that will explore the interplay between uni-directional substrate corrugation and chemical surface treatment. It may lead to viable and economical ways of obtaining BCP films with defect-free lateral alignment.

\end{abstract}

\maketitle

\section{Introduction}

Block copolymers (BCP) are a particular class of polymeric systems whose chain design is tightly connected with their functionality and applications.
The BCP chains are composed of two or more
chemically-distinct blocks that are covalently tethered together. These systems
spontaneously self-assemble into exquisitely ordered nano-structures with characteristic length scale in the range of a few nanometers to hundreds of nanometers~\cite{Bates99}.  For example, well-studied  di-BCP melts produce a rich variety of three-dimensional morphologies including lamellae, hexagonally close-packed cylinders, BCC packing of spheres,
and gyroid networks~\cite{Bates94}. The wealth of these micro-phases continues to be intensely investigated by exploring chain architecture and chemical composition~\cite{Wang95,Bates03,Hadji08,Bates12a}, or confining BCP films in finite geometries~\cite{Russell05,Li13}.

Much advance has been done in recent decades on understanding the micro-phases of BCP in bulk and in film geometries. However, technology transfer into patterning applications crucially relies on controlling of structure and local alignment~\cite{Epps13}.
Such patterning technology might replace photolithography techniques in microelectronics, magnetic storage, solar cells and novel optical devices~\cite{Kim10,Herr11}. Many techniques have been developed to attain such structural control, including chemical patterning of the substrate~\cite{Nealey03,Nealey05}, graphoepitaxy~\cite{Park07}, nanoimprint lithography~\cite{Li04,Man12}, and solvent annealing~\cite{Sinturel13}.

The orientation of anisotropic phases (lamellae, cylinders) and their lateral (inplane) ordering are two key factors in achieving structural control of BCP thin films~\cite{Nunns13}.
A perpendicular orientation with
respect to the underlying substrate of BCP lamellae or cylinders is usually desirable for many materials and engineering applications~\cite{Hamley09}. It has been implemented successfully by varying the BCP~\cite{Khanna06} and substrate characteristics~\cite{Sivaniah03,Sivaniah05,Bita08,Liu09,Kulkarni12,Char13}. However, full control of the lateral ordering in industrial processes is still an ever-challenging issue. In particular, an unsolved problem is how to produce large defect-free arrays of BCP films in an economical and viable fashion, as is required in technological applications~\cite{Bates14}.

\begin{figure}[h!t]
\begin{center}
{\includegraphics[bb=0 0 350 268, scale=0.65,draft=false]{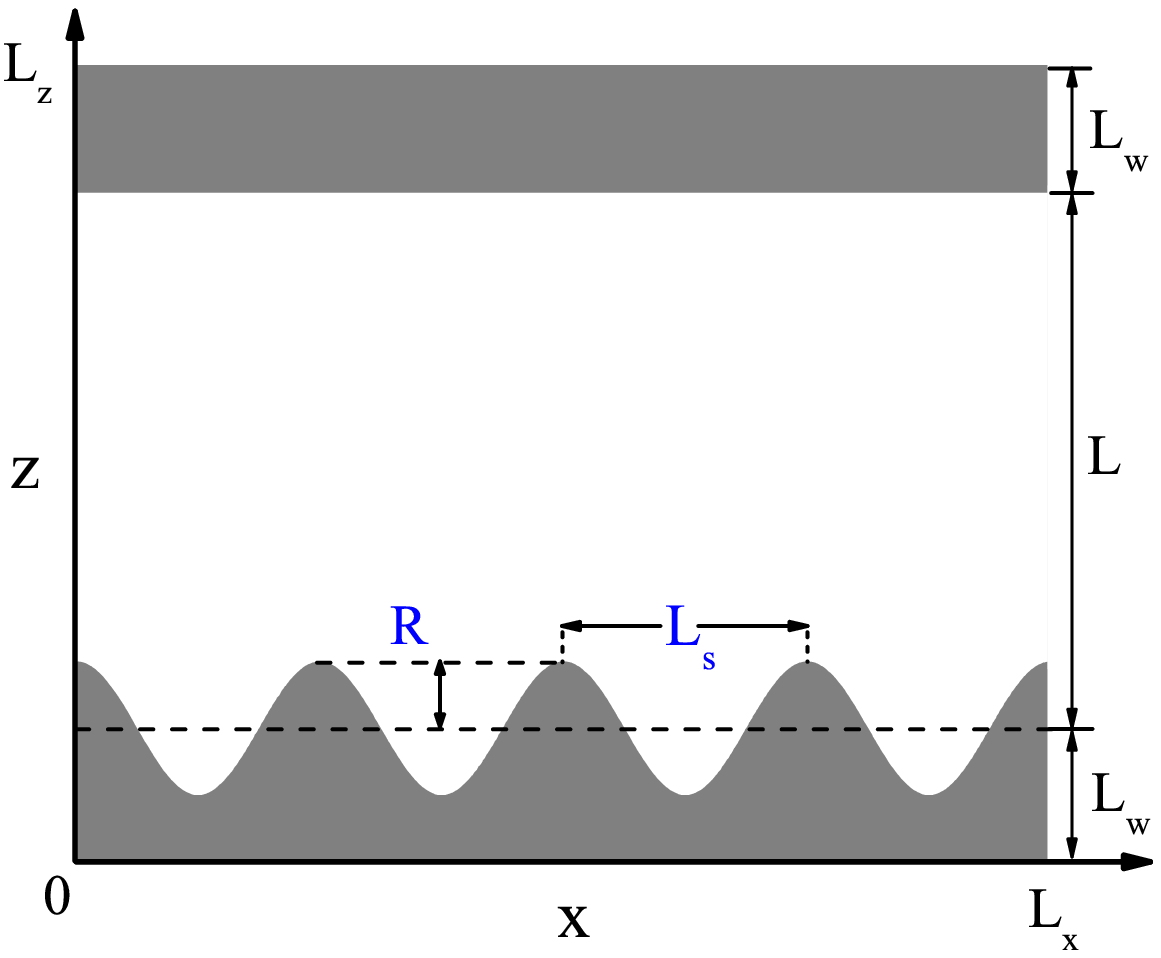}}
\caption{
\textsf{Schematic illustration of a BCP film confined between two surfaces. The 3D calculation box
is of size $L_x\times L_y \times L_z$, and the averaged BCP film thickness is $L=L_z-2L_w$, where $L_w$ is the average wall thickness within the box.
The corrugated substrate is described by a height function: $h(x,y)=R\cos
(q_s x)$, with lateral wavenumber $q_s=2\pi/L_s$ and amplitude $R$. The substrate corrugation describes
trenches that are translationally invariant in the $y$-direction. }}
\end{center}
\end{figure}
Experimental techniques to orient lamellae in a perpendicular direction with respect to the substrate are well established.  However, it is harder to control the film inplane ordering. In order to produce a defect-free and perfect inplane alignment, some experimental groups used corrugated substrates with a sinusoidal~\cite{Vega13,Aissou13}, square-waved~\cite{Park09a}, or saw-toothed~\cite{Russell09} height variation. For example, Park et al~\cite{Park09a} investigated the self-assembly of BCP thin film on square-waved substrate. They reported that when the lamellar BCP film thickness is smaller than the corrugation amplitude, the inplane alignment was found to be perpendicular to the long axis of the substrate trenches, and exhibit almost perfect ordering with no inplane defects. On the other hand, for thicker films, the substrate topography effect is weaker, resulting in a lost of the perfect inplane alignment and quite a number of defects. We note that in a recent and related (though preliminary) study~\cite{Kim14}, improved alignment of thicker lamellar forming BCP films was reported for blends of two di-BCPs having different block sizes.

It seems from the above-mentioned experimental studies~\cite{Vega13,Aissou13,Park09a,Russell09,Kim14} that, at present, there is only partial success in aligning defect-free lamellar forming BCP films as induced by substrate corrugation. On the theoretical side~\cite{Turner1992,Yoav03,Yoav05,Ranjan12,Ye14,Man15}, previous works have been mainly devoted to study the transition between parallel to perpendicular lamellar phases in presence of a substrate. Most of these studies have been done in two dimensions and cannot address the important experimental issues related to inplane alignment, which intrinsically requires a 3D study.

In the present work, we address in detail how substrate corrugation induces inplane alignment of 3D BCP films. Our findings suggest that perfect inplane alignment is achieved as a synergy of geometrical effects of uni-directional substrate trenches, together with a weak preference of the substrate towards one of the BCP components.

The outline of the paper is as follows. In the next section, we present the details of our model, and explain how the BCP structure is obtained for corrugated substrates using self-consistent field theory (SCFT). In Sec. III, we present our results, and in Sec. IV we discuss the surface-induced alignment and its relation to current and future experiments. We also present in Sec. IV our conclusions and some future prospects.

\section{Model}

Our simulation setup is shown schematically in Fig.~1. The simulations are performed in a three-dimensional (3D) box of size $L_x\times L_y\times L_z$. The technical details of using masking method to treat the boundaries can be found in Ref.~\cite{Man15}, and will not be repeated here.
The top surface is taken as a flat interface in order to
mimic the BCP surface with air, while the bottom surface is sinusoidally corrugated along the $x$-direction and
translationally invariant along the $y$-direction. The substrate height function is then written as
\begin{equation}
\label{corrugate}
h(x,y)=R\cos(q_s x)\, ,
\end{equation}
and describes periodic surface trenches having a single $q$-mode along the $x$-direction with wavenumber $q_s$, periodicity $L_s=2\pi/q_s$ and amplitude $R$. The average thickness of the BCP film, $L$, is averaged over the
corrugation of the substrate (see Fig.~1).
Although the present study is restricted to simple sinusoidal corrugations, our simulations can equally be applied to other types of surface corrugation such as saw-toothed, square-wave, etc.

Self-consistent field theory (SCFT) is used to investigate the lamellar phase of A/B di-BCP, with a natural periodicity in our thin film geometry, $L_0=2\pi/q_0$. The 3D BCP film is confined between a flat top surface and a corrugated bottom one as defined in Eq.~(\ref{corrugate}). We consider a melt of $n_c$ chains, each composed of $N=N_A+N_B$ monomers. The A-component molar fraction is $f=N_A/N$ and that of the B-component is $1-f$.
Hereafter, we concentrate on symmetric di-BCP, i.e., $f=0.5$.
All lengths are rescaled with the chain radius of gyration, $R_g=\sqrt{Nb^2/6}$, where $b$ is the Kuhn length taken for simplicity to be the same for the two blocks.

The Flory-Huggins parameter between the A and B monomers (in units of $k_BT$) is $\chi_{\rm AB}$, and $u=\chi_{{\rm sA}}-\chi_{{\rm sB}}$ is the relative interaction between the substrate and the
A/B components, where $\chi_{{\rm sA}}$ ($\chi_{{\rm sB}}$) is the interaction parameter between the substrate and the A (B) component. This choice means that a positive $u$ induces substrate preference of the A component. A detailed formulation of the numerical procedure
(pseudo-spectral method with FFTW) and its implementation to SCFT modelling of BCP systems can be found elsewhere \cite{Bosse07, Hur09, Takahashi12}.

The SCFT formulation gives the local density for the A and B components, $\phi_{\rm A}(\mathbf{r})$ and $\phi_{\rm B}(\mathbf{r})$, respectively.
There are two orientations of the BCP film with respect to the substrate, and they are shown schematically in Fig.~2. The parallel orientation is denoted $\rm{L}_\parallel$ and is shown in Fig.~2(a), while the perpendicular one is denoted as L$_\perp$. The L$_\perp$ phase can be further divided into two orientations, L$_\perp^{1}$ and L$_\perp^{2}$, which are orthogonal to each other as shown in Fig.~2(b) and (c). Whereas L$_\perp^{1}$ is orthogonal to the trench long-axis, the L$_\perp^{2}$ ordering is parallel to the trenches.  The BCP films in our study are in contact with a uni-axial corrugated substrate that has a small preference toward one of the two BCP components.
In the next section we will show under what conditions it is possible to obtain a defect-free L$_\perp$ phase.

\begin{figure*}[h!t]
\begin{center}
{\includegraphics[bb=0 60 550 202, scale=0.8,draft=false]{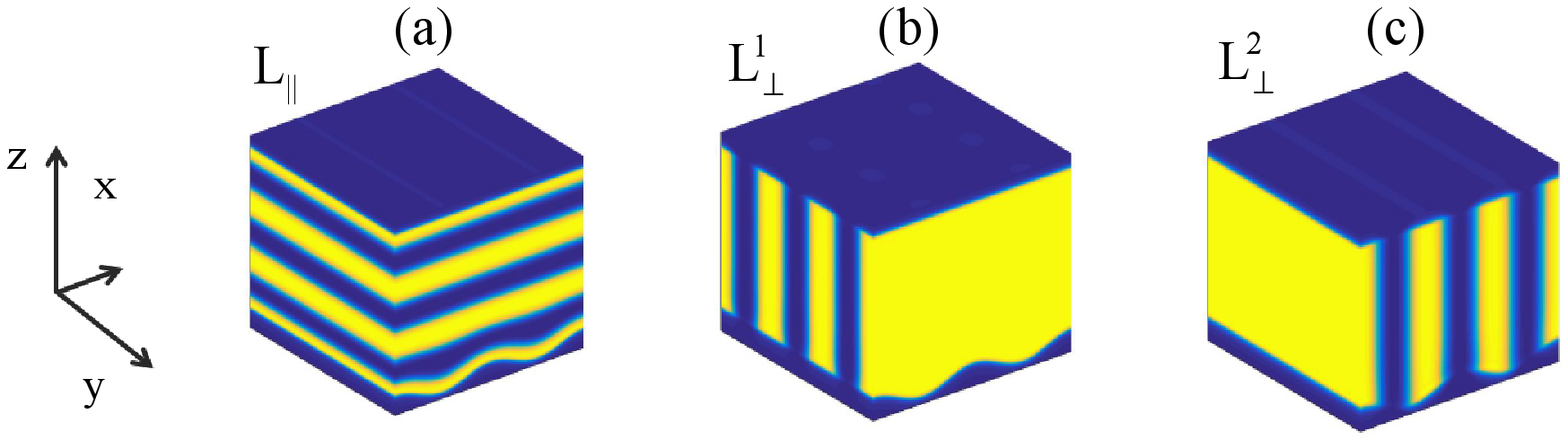}}
\caption{
\textsf{Three orthogonal lamellar orientations for BCP films in contact with
a bottom substrate that is sinusoidally corrugated. A-rich regions are denoted by yellow and B-rich by blue. In (a) the lamellar orientation is parallel to the substrate, and the phase is denoted as $\rm{L}_\parallel$, while in (b) and (c) we show the two orientations that are perpendicular to the substrate (and to each other). In (b) the lamellae  are perpendicular to the long axis of the substrate trenches and the phase is denoted as L$_\perp^{1}$. In (c) the lamellae are  parallel to the trench long-axis and the phase is denoted as L$_\perp^{2}$.
}}
\end{center}
\end{figure*}

\begin{figure*}[h!t]
\begin{center}
{\includegraphics[bb=0 86 545 335, scale=0.80,draft=false]{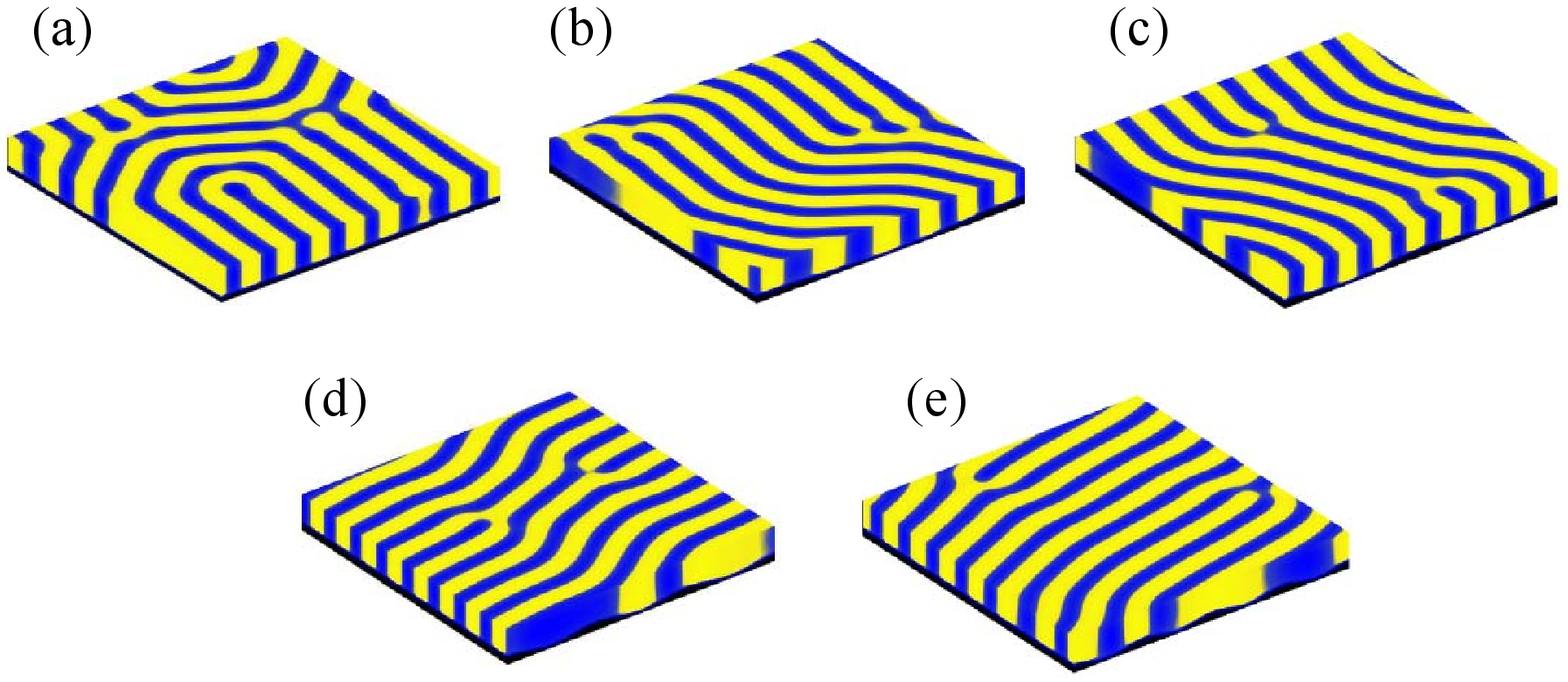}}
\caption{
\textsf{3D BCP films of thickness $L=3.0=0.9L_0$ inside a simulation box of size, $L_x=L_y=26.8=8L_0$, and $L_z=4.0=1.2L_0$. Both the top flat surface and the bottom substrate are neutral, $u=0$. In (a), the substrate is flat, while in (b)-(e) it is corrugated with $R=0.25$ and with corrugation periodicity, $L_s=26.8=8L_0$ in (b), $L_s=13.4=4L_0$ in (c), $L_s=8.9=2.7L_0$ in (d), and $L_s=6.7=2L_0$ in (e). All lengths are rescaled by the chain radius of gyration, $R_g$. In all parts (a)-(e), the resulting lamellar phase is perpendicular to the substrate and has many inplane defects.}}
\end{center}
\end{figure*}

\begin{figure*}[h!t]
\begin{center}
{\includegraphics[bb=0 115 510 330, scale=0.81,draft=false]{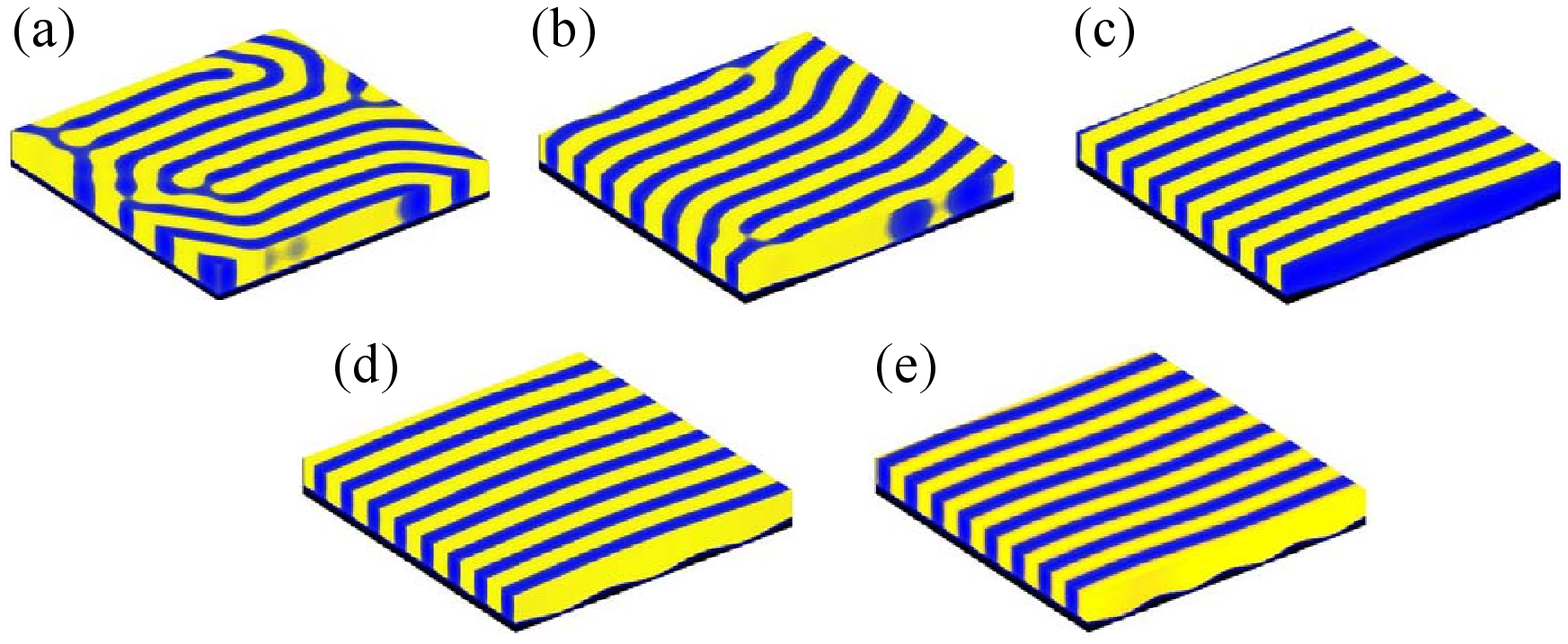}}
\caption{
\textsf{3D BCP films of thickness $L=3.0=0.9L_0$ inside the same simulation box as in Fig.~3. The top flat surface is neutral and the bottom substrate has a small preference, $u=0.5$. In (a), the substrate is flat ($R=0$), while in (b)-(e), it is corrugated with $R=0.25$. The lamellar phase has a perpendicular orientation with respect to the substrate, with inplane defects in (a) and in (b), where $L_s=8L_0$. The lamellar structure is a defect-free L$_\perp^{1}$ (orthogonal to the trench axis) in (c)-(e), with $L_s=4L_0$ in (c), $L_s=2.7L_0$ in (d), and $L_s=2L_0$ in (e).}}
\end{center}
\end{figure*}

\section{Results}

\subsection{Defect-free Thin Film of Lamellar BCP}

The structure of the BCP film is obtained via the SCFT method applied to the weak segregation, $N\chi_{AB}=12$. For this $N\chi_{AB}$ value, the BCP natural periodicity in our thin film geometry is $L_0=3.35$ (in units of $R_g$). We performed simulations at this temperature (close to ODT) because the mobility of BCP chains is enhanced and it is much easier to anneal away the inplane defects. The BCP film of thickness $L=3=0.9L_0$ is simulated inside a 3D box of cross-section $L_x\times L_y = 26.8\times 26.8=8L_0\times 8L_0$, and of height $L_z=4.0=1.2L_0$. Figure~3 shows the thermal equilibrated state of a BCP film that is casted on neutral substrate, $u=0$. In (a), the substrate is flat ($R=0$), while in (b)-(e) the substrate is corrugated with $R=0.25$, and has a decreasing periodicity:  $L_s=2\pi/q_s=26.8=8L_0$ in (b), $L_s=13.4=4L_0$ in (c), $L_s=8.9=2.7L_0$ in (d), and $L_s=6.7=2L_0$ in (e). For neutral substrates that are either flat or corrugated, the BCP film forms a L$_\perp$ phase, but has many defects with no inplane alignment. For the same neutral and corrugated substrate, we repeated the simulations with different choices of random initial conditions. Convergence is always obtained to a L$_\perp$ phase with a multitude of defects. As the initial conditions are changed, defects are always found inside the BCP film, but at different and randomly distributed positions.

In Fig.~4, we repeat the same simulations as in Fig.~3, but with a small added substrate preference, $u=0.5$ (in units of $k_B T$). Note that $u\ll N\chi_{AB}$ is chosen to be  small enough so that it does not induce the $\rm{L}_\parallel$ phase~\cite{Man15}. For the flat substrate in Fig.~4(a) and the corrugated substrate with large periodicity of $L_s=8L_0$ in 4(b), we obtain an L$_\perp$ phase with many inplane defects. As the corrugation periodicity decreases (while keeping $u$ and $R$ constant), a defect-free L$_\perp^{1}$ is found in 4(c)-4(e), where $L_s$ varies between $4L_0$ to $2L_0$.

Two conclusive remarks, supported by Fig.~4, can be made at this stage. (i) In order to get a perfect (defect-free) L$_\perp$ phase, the substrate corrugation periodicity should lie in the range of $2L_0\le L_s \le 4L_0$ when the other set-up parameters are: $u=0.5$, $R=0.25\simeq 0.07L_0$ and $L=0.9L_0$. (ii) The stability of the defect-free L$_\perp$ crucially depends on a moderate surface preference, parameterized by a small $u$ value of the corrugated substrate.

The defect-free L$_\perp$ phase can be reproduced in an entire range of the corrugation amplitude, $R$, and
periodicity, $L_s$, as long as the substrate exhibits a small preference towards one of the BCP
components and its corrugation is unidirectional. An example of a perfectly aligned L$_\perp^{1}$ is shown in Fig.~5 for $L_s=4L_0$ and $u=0.5$. Further checks showing similar behavior were also performed for $L_s=2L_0$ without presenting them here explicitly. The L$_\perp$ phase shown in Fig.~5 is defect-free for the same range in corrugation amplitude, varying between $R=0.15$ in 2(a) to 0.45 in 2(c).

From the presented results, a prediction with experimental consequences emerges. It suggests that the perfect L$_\perp$ alignment is induced by the substrate roughness, $R\geq0.15$ ($\approx0.05L_0$), but is not sensitive to the specific value of $R$.
The defect-free L$_\perp$ structure can be reproducible and was repeatedly obtained for several different random initial conditions of the BCP film, always in contact with the same corrugated substrate.

\begin{figure*}[h!t]
\begin{center}
{\includegraphics[bb=0 0 525 120, scale=0.78,draft=false]{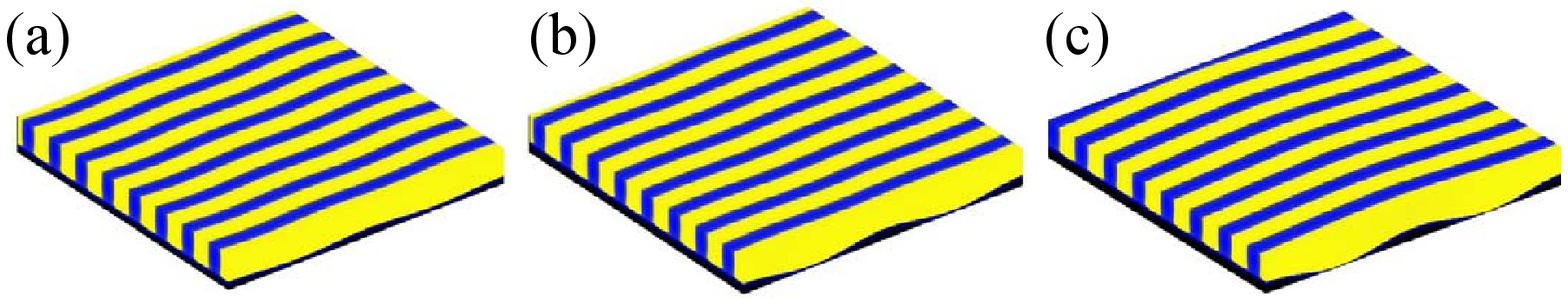}}
\caption{
\textsf{The effect of the substrate roughness amplitude, $R$, on the 3D BCP film structure. In (a), (b) and (c), the amplitude $R$ is $0.15,\, 0.35$ and $0.45$, respectively.
The film and simulation box are the same as in Fig.~4. In all figure parts, the substrate is corrugated with periodicity $L_s=4L_0$, and the resulting lamellar phase in all cases is defect-free L$_\perp^{1}$.
}}
\end{center}
\end{figure*}

\subsection{Conditions for the stability of the Perpendicular Phase}

\begin{figure}[h!t]
\begin{center}
{\includegraphics[bb=0 0 360 280, scale=0.65,draft=false]{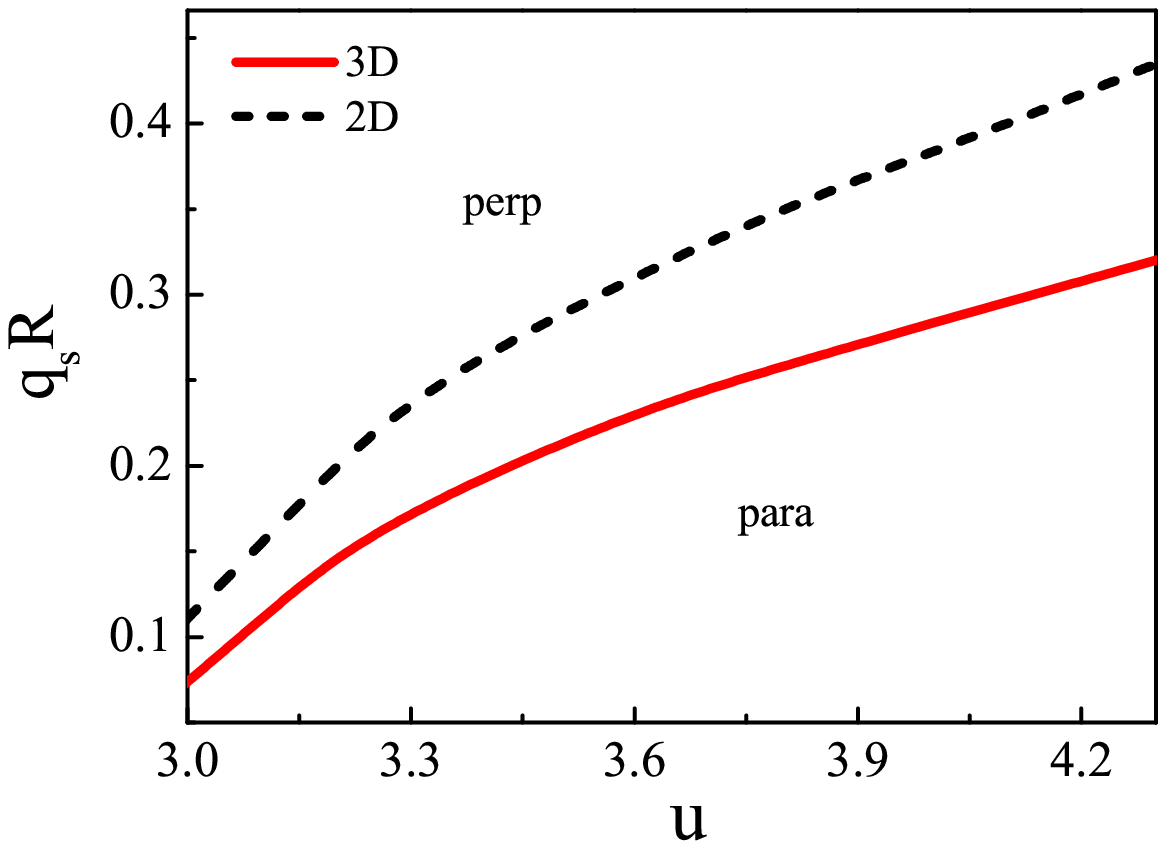}}
\caption{
\textsf{The parallel ($\rm{L}_\parallel$) to perpendicular (L$_\perp$) phase transition in the ($u,q_sR$) plane. A comparison is shown between a 3D (red solid line) and a 2D (dashed black line) system. In 3D, the perpendicular phase is the L$_\perp^{1}$, and has an extended region of stability as compared to 2D. The top flat surface is neutral and $N\chi_{AB}=25$.
}}
\end{center}
\end{figure}

\begin{figure}[h!t]
\begin{center}
{\includegraphics[bb=0 0 380 280, scale=0.65,draft=false]{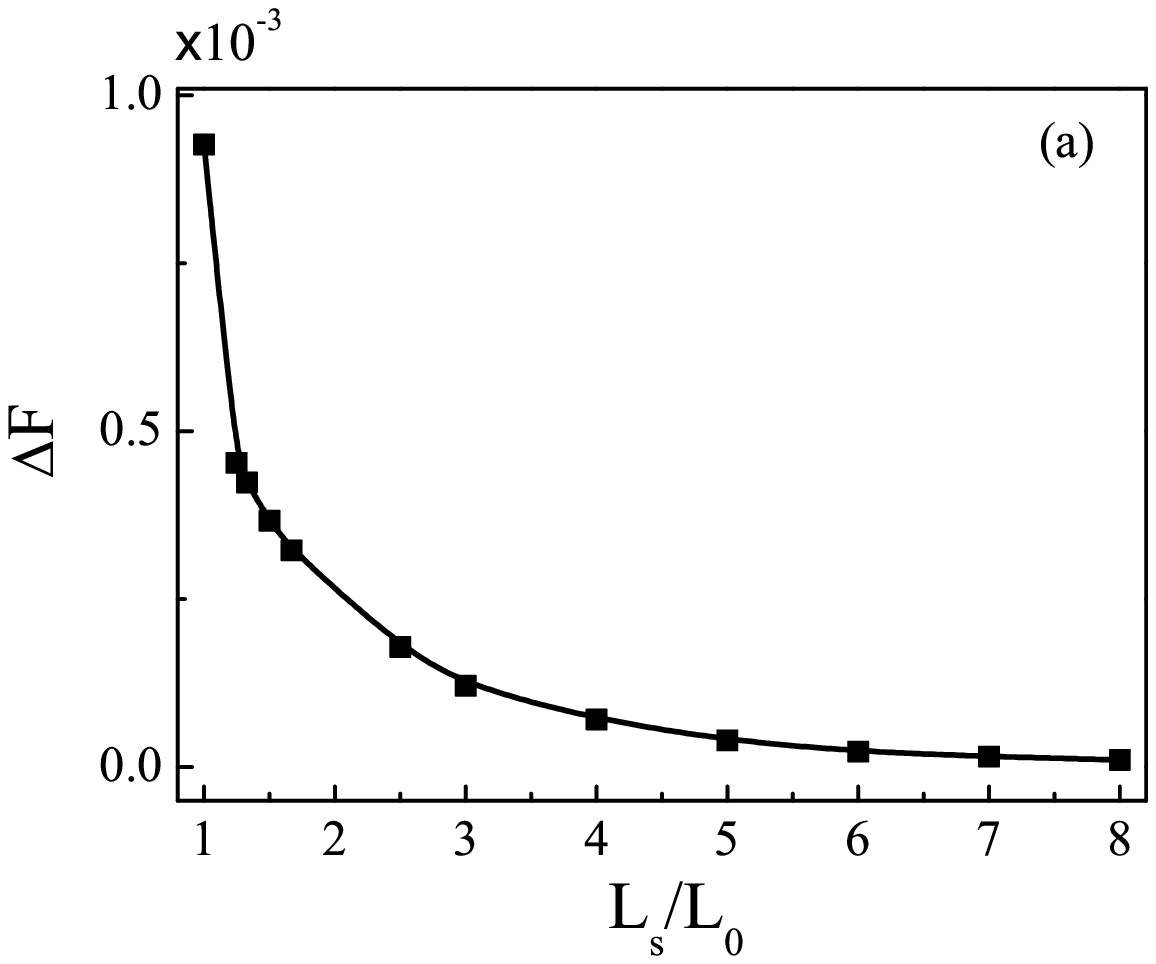}}
{\includegraphics[bb=0 0 380 280, scale=0.655,draft=false]{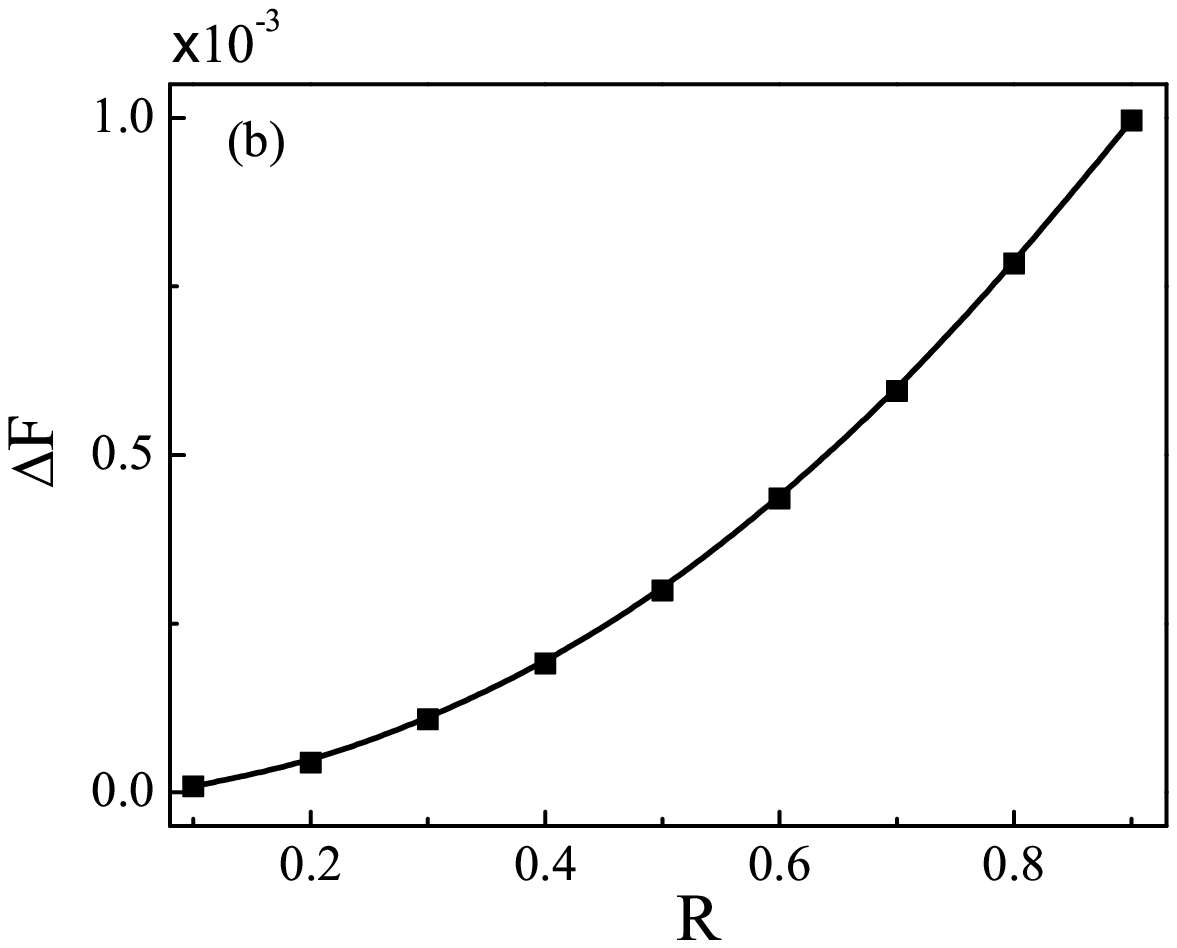}}
\caption{
\textsf{The free-energy difference per unit volume, $\Delta F=F_\perp^{2}-F_\perp^{1}$, in units of $k_B T$ is shown to be: (a) a decreasing function of $L_s/L_0$, and (b) an increasing  function of $R$. All other parameters are as in Fig.~4. }}
\end{center}
\end{figure}

In order to understand why we obtain defect-free L$_\perp$ phases, the relative stability of lamellar phases with the three orthogonal orientations ($\rm{L}_\parallel$, L$_\perp^{1}$ and L$_\perp^{2}$) is studied by comparing their free energies.
The effect of surface preference is addressed more quantitatively in Fig.~6, where the transition between L$_\perp$ and $\rm{L}_\parallel$ is shown in the $(u, q_sR)$ plane. We use larger $u$ values (between 3.0 and 4.2) in order to compare the present 3D study (red line) with the previous two-dimensional (2D)(dashed black line) study of Ref.~\cite{Man15}, where the BCP was taken in the strong segregation regime, $N\chi_{AB}=25$. As can be seen in Fig.~6, the region of stable L$_\perp$ is quite extended, while  the stabilization of the $\rm{L}_\parallel$ phase requires large $u$ values. For the weaker segregation regime, such as $N\chi_{AB}=12$ that is used in most of the present study, the stability of the L$_\perp$ is even further enhanced.

Note that in 2D there is only one perpendicular phase to the substrate, while in 3D there are two perpendicular orientations: the L$_\perp^{1}$ phase that is orthogonal to the trenches, and the L$_\perp^{2}$ whose orientation is parallel to the trenches (see Fig.~2). With infinite trenches in the $y$-direction,
the latter L$_\perp^2$ phase in 3D resembles the L$_\perp$ phase in 2D.
As will be shown below, the L$_\perp^{1}$ is  more stable than L$_\perp^{2}$, resulting in a larger region in Fig.~6 occupied by the stable 3D perpendicular phase in the phase diagram, as compared with the 2D case.

In Fig.~7, we show the relative stability of the two perpendicular orientations, L$_\perp^{1}$
and L$_\perp^{2}$, for a fixed and relatively small value of the substrate preference, $u=0.5$. The free-energy difference per unit volume, $\Delta F= F_\perp^{2}-F_\perp^{1}$ (in units of $k_B T$),
is plotted as function of the periodicity ratio, $L_s/L_0$ in (a) and as function of the substrate
amplitude $R$ in (b). Since we are in the weak segregation regime ($N\chi_{AB}=12$), the free-energy difference between the two orientations is small, of order 10$^{-3}$ (in units of $k_B T$ per unit volume). As $\Delta F$ is positive for the entire range of parameters, the L$_\perp^{1}$ phase (i.e., perpendicular to the trenches) is the most stable one. In the limit of $R\to 0$ and $L_s/L_0\to \infty $, the corrugated substrate approaches a flat one and $\Delta F$ becomes vanishingly small.

\begin{figure}[h!t]
\begin{center}
{\includegraphics[bb=0 0 380 280, scale=0.65,draft=false]{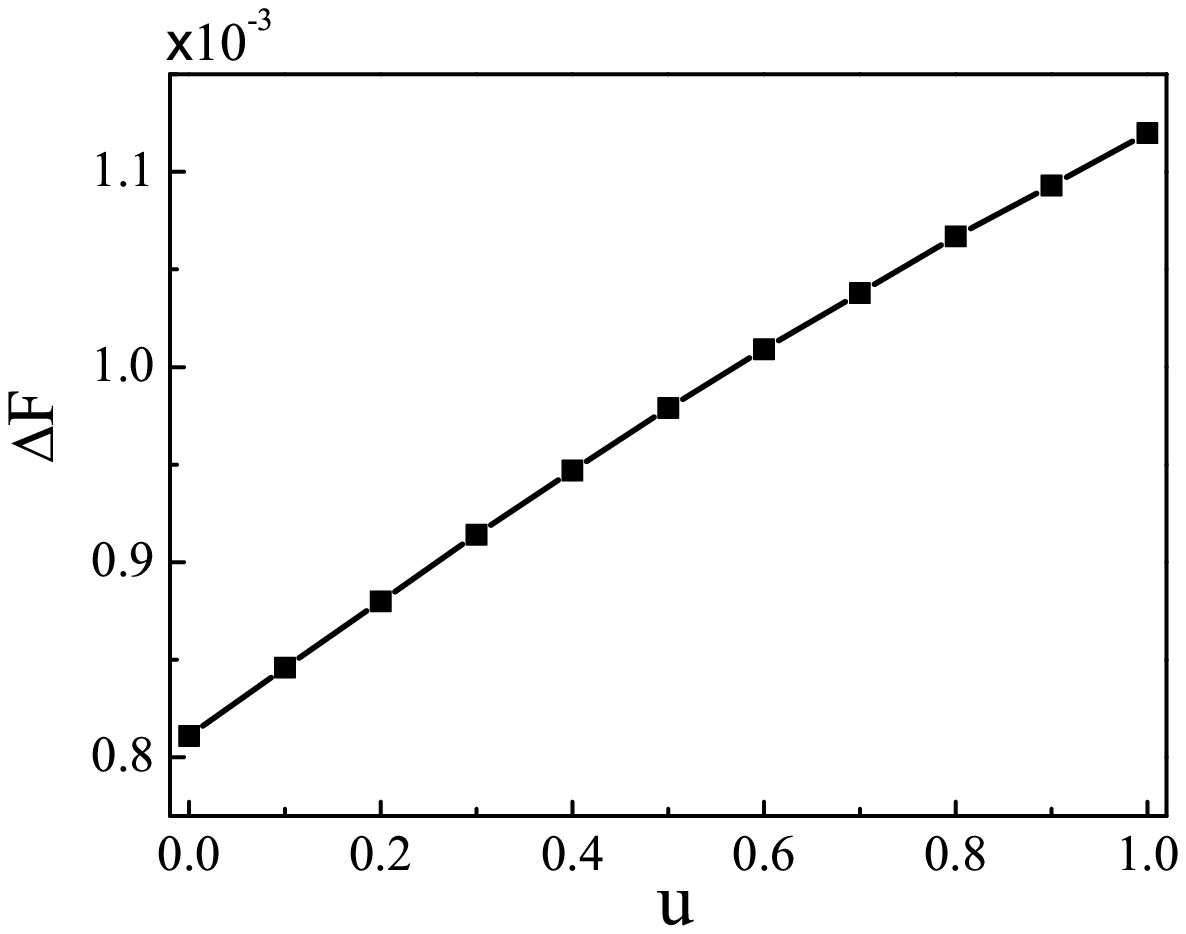}}
\caption{
\textsf{The dependence of $\Delta F$ (in units of $k_B T$ per unit volume) on $u$ showing an increase of $\Delta F$ with $u$. This increase implies that the surface preference enhances the stability of the L$_\perp^{1}$ phase. All other parameters are the same as in Fig.~4.}}
\end{center}
\end{figure}

As the substrate preference is found to play an important role in stabilizing the L$_\perp^{1}$ phase, we explore in Fig.~8 the role of the preference field $u$ on the free-energy difference, $\Delta F$, for a fixed $L_s=4L_0$ and $R=0.45$.
As can be seen in Fig.~8, $\Delta F$ increases with $u$ making L$_\perp^{1}$ more stable as compared to L$_\perp^{2}$.

\begin{figure}[h!t]
\begin{center}
{\includegraphics[bb=0 50 180 280, scale=0.95,draft=false]{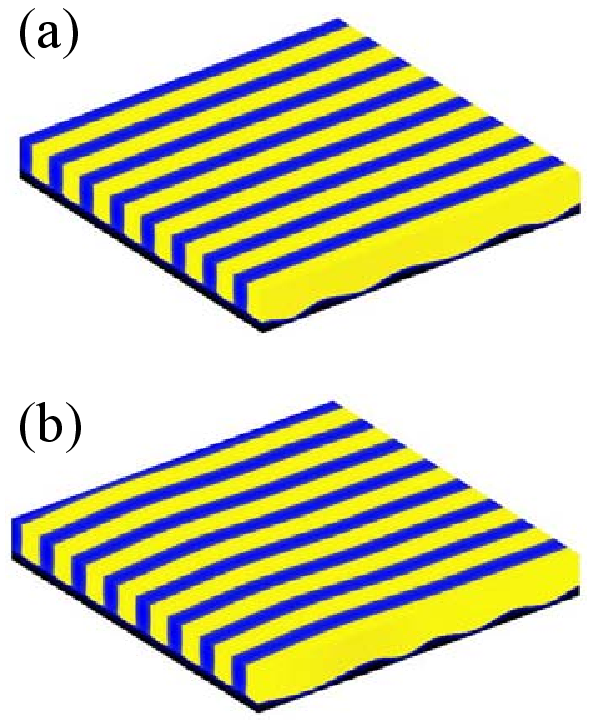}}
\caption{
\textsf{3D BCP film in contact with a multi-mode sinusoidal substrate. The film and simulation box are the same as in Fig. 4. The amplitude of the corrugated substrate is $R=0.25$, while the corrugation periodicity is a combination of four different periodicities. In (a) the four periodicities are: $1.5L_0$, $2.0L_0$, $2.8L_0$ and $1.7L_0$, and in (b) they are: $3.0L_0$, $1.5L_0$, $1.9L_0$ and $1.6L_0$. Both lamellar phases are defect-free L$_\perp^{1}$.
}}
\end{center}
\end{figure}

\subsection{Multi-mode Corrugated Substrates}

So far, the corrugated substrate was characterized by a single uni-directional sinusoidal mode for the entire simulation box. In order to model the experimental situation where substrates have an entire distribution of $q$-modes, we present in Fig.~9 an example of a BCP film in contact with a  multi-mode corrugated substrate. As our simulation box has the size of about eight lamellar periodicities, $L_x=L_y\simeq 8L_0$,
we cannot simulate a real multi-mode random substrate. Instead, we use a substrate composed of four sections, each with a different periodicity between $1.5L_0$ and $3.0L_0$ (see Fig.~9). An important point to consider is that the substrate uni-directionality is strictly observed, as well as the weak value of its preference parameter, $u\gtrsim 0$.
Keeping these points in mind, Fig.~9 indeed shows that the BCP alignment is not sensitive to such a choice of a multi-mode substrate. The obtained BCP phase is still a defect-free L$_\perp^{1}$.

\section{Discussion and Conclusions} 

In this paper we proposed a direct self-assembly scheme to obtain defect-free perpendicular BCP lamellar phases by manipulating both the geometrical roughness and chemical preference of the underlying corrugated substrate.
Our results suggest that there is an enhanced synergy between these two important substrate properties, as long as we maintain uni-directional substrate corrugations and a small surface preference.
However, the two surface perturbations cannot be too large. If a large chemical preference is introduced, it will destabilize the L$_\perp$ phase (and stabilize the parallel phase, $\rm{L}_\parallel$). Similarly, a large $q_sR$ values will cause pronounced deformations of the L$_\perp$, with potential defects close to the corrugated substrate~\cite{Man15}.

In terms of the defect-free L$_\perp$ phase, our results are robust as can be seen in Figs.~3-5. Changes in the periodicity $L_s$ and amplitude $R$ of the substrate corrugation do not affect the alignment, as long as the uni-directional corrugation is moderate and the substrate preference parameter is small. Furthermore, the defect-free L$_\perp$ phase is obtained even for multi-mode corrugated substrates as shown in Fig.~9. Another reason for the claimed robustness is that the defect-free L$_\perp$ phase is obtained for different choices of the random initial condition of the film structure, mimicking a film preparation above the ODT or a disordered BCP solution that is spin-casted onto the substrate.

As in any numerical scheme, we are limited by the system  finite size. To check our predictions, we also simulated (without showing results) larger lateral system sizes of about $10L_0$, while the film thickness was kept the same as before, $L=0.9L_0$. As long as $2L_0\leq L_s\leq4L_0$, we still obtain the perfect L$_\perp$ phase. Although our simulation results are indicative, they are not completely sufficient to predict that a similar combination of substrate topography and chemical preference can induce defect-free L$_\perp$ for systems with much larger lateral size, as is required in experiments. This is a deficiency of all simulation schemes as the simulation box has a finite size that is often smaller than the experimental relevant sizes. In addition, as experiments have reported a dependence of the film orientation on its thickness~\cite{Bong09}, we conjecture that a perfect L$_\perp$ phase of different thicknesses or sizes requires different values of the roughness parameters ($q_s$ and $R$), and preference parameter, $u$. This conjecture has yet to be confirmed in future studies.

In a fraction of our simulations, instead of obtaining a defect-free L$_\perp^1$ phase, the system converged towards a tilted L$_\perp$ phase. These tilted L$_\perp$ lamellae have a perfect perpendicular orientation with respect to the substrate and are aligned with a small tilt angle with respect to the inplane direction of the L$_\perp^1$ phase, but still without any inplane defects. We believe that such a tilt angle results from the finite system size in the simulations, given the influence of the vertical boundaries of the simulation box on the orientation of the L$_\perp$ phase. As such tilted L$_\perp$ phases still maintain a perfect lateral ordering, their existence will not affect the implications of our findings toward applications of thin BCP films in patterning technologies.

The tendency of the L$_\perp$ phase to align in the L$_\perp^1$  direction that is orthogonal to the trenches (as compared to the parallel one, L$_\perp^2$) is shown in Figs.~7 and 8. From the figures it is observed that the free energy difference $\Delta F=F_\perp^2-F_\perp^1$ is always positive. This observation is consistent with previous theoretical findings and experiments~\cite{Park09a,Kim14}. Pickett et al~\cite{Pickett93} proposed an explanation of why the L$_\perp^{1}$ should be more stable than L$_\perp^{2}$ for most BCP films and surface parameters. According to their explanation, when the BCP chains are in contact with a corrugated substrate, the chain deformations are very different in the two L$_\perp$ orientations. For the L$_\perp^{1}$, the chains are oriented along the ``smooth" surface direction and they are almost not deformed, while for the  L$_\perp^{2}$ phase, the chains follow the contour of the substrate corrugation. This means that for the latter  L$_\perp^{2}$ phase, the chains are locally extended or compressed, which cost more energy.

Furthermore, as can also be seen in Figs.~7 and 8, increasing the substrate roughness (decreasing $L_s\sim q_s^{-1}$ in Fig.~7) or substrate preference (increasing $u$ in Fig.~8) causes $\Delta F$ to increase. This means that the L$_\perp^{1}$ phase becomes more stable than the L$_\perp^{2}$ phase as function of $q_sR$ and $u$. This tendency can be understood in terms of local extension or compression of the chains that is more pronounced in the L$_\perp^{2}$ phase than in the L$_\perp^{1}$ phase.  For example, when the substrate preference towards the A component increases, the preferred A-rich domain close to the substrate will expand. As the chains follow the contour of the substrate corrugation in the L$_\perp^{2}$ phase but it is parallel in L$_\perp^{1}$ phase, the cost energy of chain extension or compression increases more in the L$_\perp^{2}$ phase than for L$_\perp^{1}$. Therefore, increasing substrate roughness and preference simultaneously has a synergistic effect, which facilitates the stability of inplane defect-free perpendicular lamellar phases.

The advantage of the weak segregation regime (small $N\chi_{\rm AB}$) of the BCP film is that it avoids getting trapped in a metastable structure with defects, because the mobility of the BCP chain is enhanced when the temperature is close to the ODT. After equilibration, one can use thermal annealing to lower the temperature towards the desired strong segregation regime, while maintaining a perfect alignment.

An interesting direction of extending the present study is to consider BCP blends. In experiments~\cite{Kim14}, such blends casted on rough substrates have been shown to enhanced the ordering of defect-free phases. We hope to further explore such films composed of BCP blends, given their great value in applications.

To the best of our knowledge, using the synergy between uni-directional corrugated substrates and a small preference field to produce defect-free phases has not yet been tested in experiments. Future experimental verifications of these scenarios may bring us closer to producing large arrays of defect-free BCP structures, while employing simple and viable techniques.

\bigskip
{\bf Acknowledgement.}~~
We would like to thank M. M$\rm{\ddot{u}}$ller for useful discussions and T. Russell for numerous suggestions and clarifications on feasible experimental setups. This work was supported in
part by grants No.~21404003 and 21434001 of the National Natural Science Foundation of China (NSFC),
and the joint NSFC-ISF Research Program, jointly funded by the NSFC under grant No.~51561145002 and the Israel Science Foundation (ISF) under grant No.~885/15.
D.A. acknowledges the hospitality of the KITPC and ITP, Beijing, China, and partial support from
the Israel Science Foundation (ISF) under Grant No.~438/12,
the United States--Israel Binational Science Foundation (BSF) under Grant No.~2012/060,
and a CAS President's International Fellowship
Initiative (PIFI).

\newpage

%

\end{document}